\documentclass[12pt]{article}

\usepackage{graphicx,subfigure,amsmath,latexsym,amssymb,times}
\usepackage{float,epsfig,times,verbatim,multirow}
\usepackage{bbold,rotating,wrapfig,color}
\usepackage{longtable,booktabs}
\usepackage{hyperref}
\hypersetup{unicode=true,
            pdftitle={Mining the Daschle Collection},
            pdfauthor={Damon Bayer, Semhar Michael},
            pdfborder={0 0 0},
            breaklinks=true}
            
\RequirePackage[round]{natbib}
\makeatletter
\renewcommand\@biblabel[1]{}
\makeatother
\evensidemargin \oddsidemargin

\def\bSig\mathbf{\Sigma}

\numberwithin{equation}{section}

\begin{document}
\markboth{Bayer and Michael}{Exploring the Daschle Collection using Text Mining Methods}
\thispagestyle{empty}
\vspace{-1in}
\title{Exploring the Daschle Collection using Text Mining Methods}
\author{{Damon Bayer and Semhar Michael}
\thanks{{Damon Bayer is a graduate student in the Department Mathematics and Statistics at South Dakota State University, Brookings, SD 57007.}; Semhar Michael is an Assistant Professor in the Department of Mathematics and Statistics at South Dakota State University, Brookings, SD 57007, email: semhar.michael@sdstate.edu.}}

\date{Received: date / Accepted: date}

\maketitle

\begin{abstract}
A U.S. Senator from South Dakota donated documents that were accumulated during his service as a house representative and senator to be housed at the Bridges library at South Dakota State University. This project investigated the utility of quantitative statistical methods to explore some portions of this vast document collection. The available scanned documents and emails from constituents are analyzed using natural language processing methods including the Latent Dirichlet Allocation (LDA) model. This model identified major topics being discussed in a given collection of documents. Important events and popular issues from the Senator Daschle’s career are reflected in the changing topics from the model. These quantitative statistical methods provide a summary of the massive amount of text without requiring significant human effort or time and can be applied to similar collections.

\end{abstract}

\normalsize

\section{Background}
\label{sect.intro}
Senator Thomas A. Daschle represented South Dakota in the United States congress for 26 years during which he served in the the House of Representatives (1978 - 1986) and the Senate (1987 - 2004). While in the Senate, Daschle served as the leader of the Democratic party from 1994-2004. The ``Senator Thomas A. Daschle Congressional Research Study" \citep{daschle17} at South Dakota State University houses a collection of of more than 2,000 linear feet of documents from Daschle's career in congress. Among these are voting records, speeches, sponsored legislation, personal papers, as well as research documents relating to special interest to the Senator, such as the effects of Agent Orange on veterans and services on American Indian reservations. 

In addition to these physical items, the collection also includes over 12,000 emails from constituents which were sent to the Senator's office from 2002 to 2004. Through this analysis, we examine this massive collection of text documents using topic modeling and a variety of other exploratory techniques in order to summarize the issues important to South Dakotans and the Senator during his career. These methods are advantageous to traditional text techniques because they allow valuable insights to be made without manual reading of 10,000's of documents, which may be time consuming or pose a privacy concern in the case of constituent emails.

We perform topic modeling on the Daschle Collection after  pre-processing and initial summary analysis. In general, topic modeling is a statistical text mining tool for automatically identifying hidden patterns in a given collection of documents and assigning them to one or more topics. Following this procedure, a human can interpret the topics to give them a more precise and understandable label. In contrast to reading or skimming documents individually, this process reduces the human effort required to sort a collection to only interpreting a small number of computer-generated topics. Therefore, it becomes an effective tool for summarizing a massive collection of documents. Topic modeling has recently been used in the humanities to explore civil war era newspaper archives \citep{nelson11}, 18th century diaries \citep{blevins11}, and even as an initial step in literary criticisms \citep{buurma15}.

The incredible size and scope of the Daschle Collection makes it an ideal candidate for implementation of statistical modeling approaches to extract important features and information. Though much of the collection currently exists solely in paper form, 40GB of data, consisting of over 4000 scanned pages of speeches, legislation, financial information, and various other documents, is already available for pre-processing and analysis. Optical character recognition techniques as ones recently developed by \citep{chaudhurietal17} will be used to convert these scanned documents into text files. These text documents will then be converted into numeric representations such as term document matrix or N-gram representation to be used for  further analysis. After initial exploratory analysis we employ the probabilistic methodology Latent Dirichlet Allocation model \citep{bleietal03}. This probabilistic modeling approach allows us to extend the methodology to a new dataset without additional human efforts in processing the data. The currently digitized set of the Daschle collection represents only a small part of the total collection. As more of the collection becomes digitized, the new documents can easily be incorporated into the analysis.

This paper develops a workflow and investigate the two types of text data in this collection. In the first case, we apply optical character recognition (OCR) methods to read in scanned documents and convert them in to text. We then implement some common algorithms to clean and perform exploratory analysis. In the second case, we explore the emails that the senator received from his constituents. Since this was already in text form, we performed cleaning and analysis directly. 

The paper is organized as follows: Section~\ref{sect.meth} presents some necessary preliminaries that are necessary to perform the analysis of the two types of data. The analysis of the two datasets is examined in Section~\ref{sect.appl}. The results of the analysis provide an insight to the vast collection. The paper concludes with a brief discussion and future work in Section~\ref{sect.disc}.

\section{Methods}
\label{sect.meth}

This section starts with a discussion of necessary preliminaries on text data mining methods. Then, focuses the Latent Dirichlet Allocation (LDA) model and its parameter estimation methods. It will also discuss the extensions of this model to incorporate time varying topic models.

\subsection{Exploratory methods}
\label{sect.expl}
\subsubsection{Data cleaning methods}
The primary method of text representation used in this paper is the bag-of-words model. For our data, one scanned document or a single email are  considered to be a single document. This model represents a
document as a list of unique words found in the document paired with the count of each word. Some modifications can be done to this process to make the data more usable. Generally this includes removing
common words or ``stop words", and ``stemming". Stop words such as ``an", ``for", and ``the" typically do not carry much meaning for some analysis and can be eliminated to simplify the matrix. Stemming can also be applied to reduce a word to its root form. For most purposes, ``fight", ``fights", ``fighting", and ``fighter" provide nearly the same information in a sentence. Stemming each of these terms to be ``fight" simplifies the matrix and accounts for their similarities.

A language model is a model that assigns a probability to a sequences of words \cite{jurafskyandmartin09}. Language models are useful in many natural language tasks, such as spelling correction. A language model would enable a computer to correct the sequence ``every night I dream of world piece" to ``every night I dream of world peace", even though “piece” and “peace” are both valid English words. Language models are also used in speech recognition, handwriting recognition, machine translation, and will be used in one of our classification methods.

In linguistics, an individual word is called a unigram. Two consecutive words are called a bigram, three consecutive words are called a trigram, etc. These n-grams can be used to construct a ``bag-of-n-grams" model. This enables us to capture the order of the words in the data, but can be problematic for smaller datasets, as each n-gram will occur less frequently than a given unigram. 

\subsubsection{Word frequency}
Word clouds and bar charts are one way to summarize text data based on word frequency. To create word clouds and bar charts for frequent words, we first removed stop words \cite{meyeretal08} and common words among all emails (``dear", ``sincerely", ``regards") or others in the documents. The resulting bags-of-words were used for input in the {\bf wordcloud} function from the wordcloud package \citep{ian14} in the statistical software R \citep{R16}. This package prints the most frequent words sized proportionally according to their frequency in a square plot.

Additionally, bar charts were constructed from the a fixed number of most frequent terms from the bags-of-words. In order to better understand which words best represented each class, we converted the frequency counts into relative frequencies. We calculated these
relative frequencies for data in each topic  and for the data outside of that class. This can be thought of as calculating the unigram probability for a word in a class, $\frac{\# word_i|topic_k}{\sum{k=1}^K \# word_1|topic_k}$ and the unigram probability for a word outside of a class, $\frac{\# word_i|topic_k^\prime}{\sum{k=1}^K \# word_1|topic_k}$. Computing the differences between the unigram probabilities in the different groups yielded a measure of each word's importance to the topic.
\subsection{Topic modeling}
\label{sect.tm1}

Topic model is a probabilistic model which aims to identify ``hidden" topics in a corpus and assign words or documents to these topics. Latent Dirichlet allocation (LDA), first proposed by Blei and his colleagues \citep{bleietal03} is the most common method of topic modeling and the one employed in this analysis. It is a generative statistical model with the following process:

For a corpus $D$, with $M$ documents with $N$ words and $K$ topics where $i \in\{1, \ldots, M\}$ indicates a specific document, $k \in\{1, \ldots, K\}$ indicates a specific topic, and $j \in\{1, \ldots, N_i\}$ indicates a specific word.
\begin{itemize}
\item[1.]  Choose $\beta_k \sim \text{Dirichlet}(\delta)$    
\item[2.]  Choose $\theta_w \sim \text{Dirichlet}(\alpha)$   
\item[3.]  For each of the $N$ words $w_i$   
\begin{itemize}
	\item[a.] Choose a topic $z_{i,j} \sim \text{Multinomial}(\theta)$    
    \item[b.] Choose a word $w_{i,j} \sim \text{Multinomial}(z_{i,j})$    
\end{itemize}
\end{itemize}

Following this model, we can view a document as a mixture of topics, with each topic being a mixture of words. In our case, the parameters of these distributions estimated using variational expectation-maximization (VEM) algorithm (see \cite{bleietal03} for details).
To find the optimal number of topics to model in our dataset, we used a
variety of metrics  proposed by \cite{griffithsandsteyvers04}, \cite{deveaudetal14}, \cite{caoetal09}, and \cite{arunetal10} as implemented in \cite{nikita16}.

\section{Data analysis}
\label{sect.appl}
In this section, we discuss the steps taken to analyze both forms of documents- paper and email. Later we present the results obtained from both data. 
\subsection{Paper documents}

While the total Daschle collection consists of over 2000 linear feet of
materials, distributed among 750 boxes, only 4,165 documents (8,034
pages) of text documents have thus far been digitized. Our analysis
focuses on this smaller subset of data, which were digitized for use in
prior research projects. As such, we expect the number of topics present to be a subset of the total represented in the complete collection.

Analysis began with applying optical character recognition (OCR) to the
documents. OCR aims to extract the text from the scanned images of the
paper documents in the Daschle collection. We used the open-source
software Tesseract (Smith 2007) implemented in the {\sc tesseract} R package \citep{ooms17a} to accomplish this task. To assess the quality of these documents we used the the open-source Hunspell \citep{hunspell19} dictionary implemented in the {\sc hunspell} R package \citep{ooms17b}. We deemed a document to be low-quality if fewer than half of the ``words'' output from the OCR software were not found in the Hunspell English dictionary. These low-quality documents were removed from the dataset and not analyzed (1,087 documents, or 26\% of the total documents). We deemed a document to be high-quality if greater than 90\% of the ``words'' output from the OCR software were found in the Hunsepll English dictionary.

\begin{figure}[ht]
 \centering
  \mbox{
       \subfigure[Low quality]{\scalebox{0.05}
      {\includegraphics{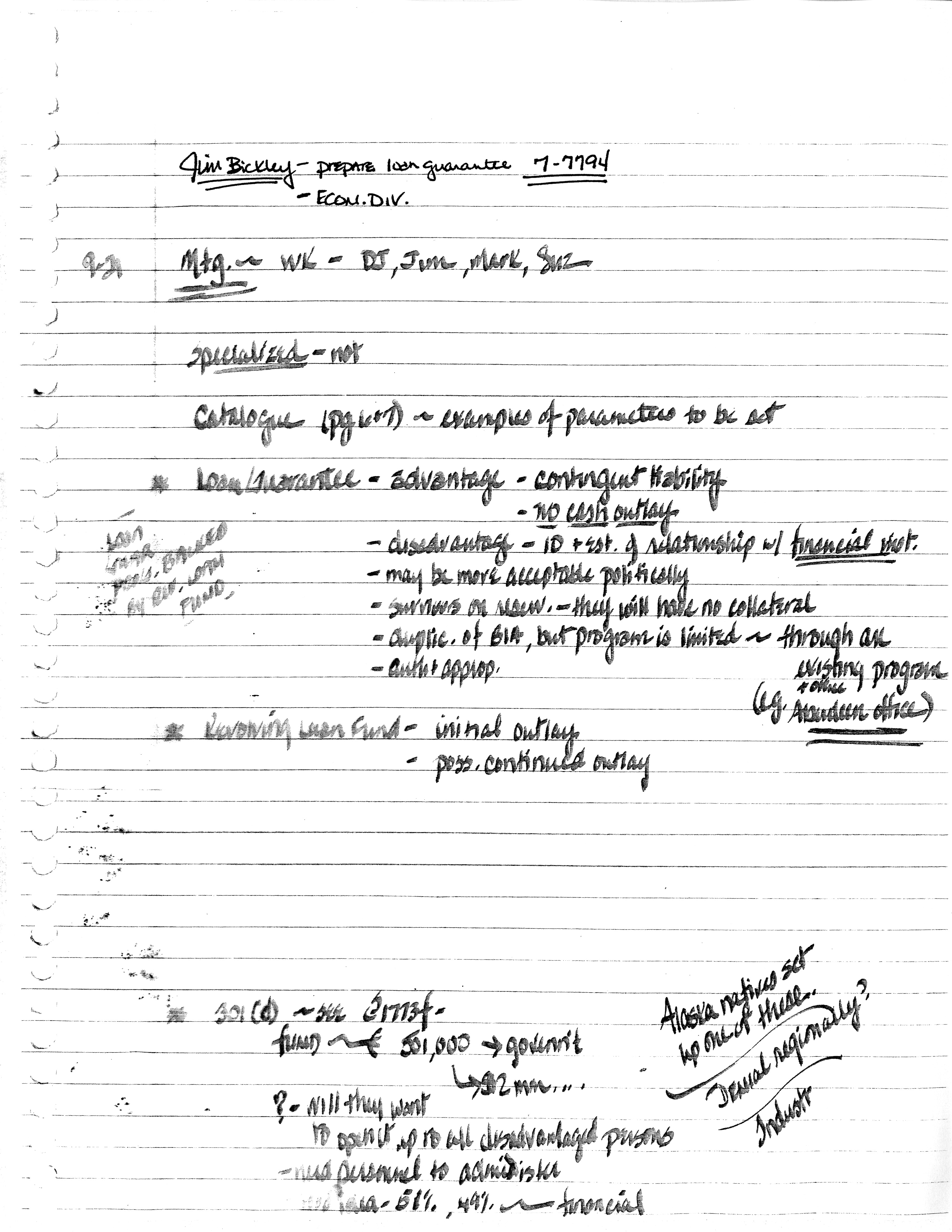}}}
}
\centering
 \mbox{
 \subfigure[High quality]{\scalebox{0.3}
      {\includegraphics{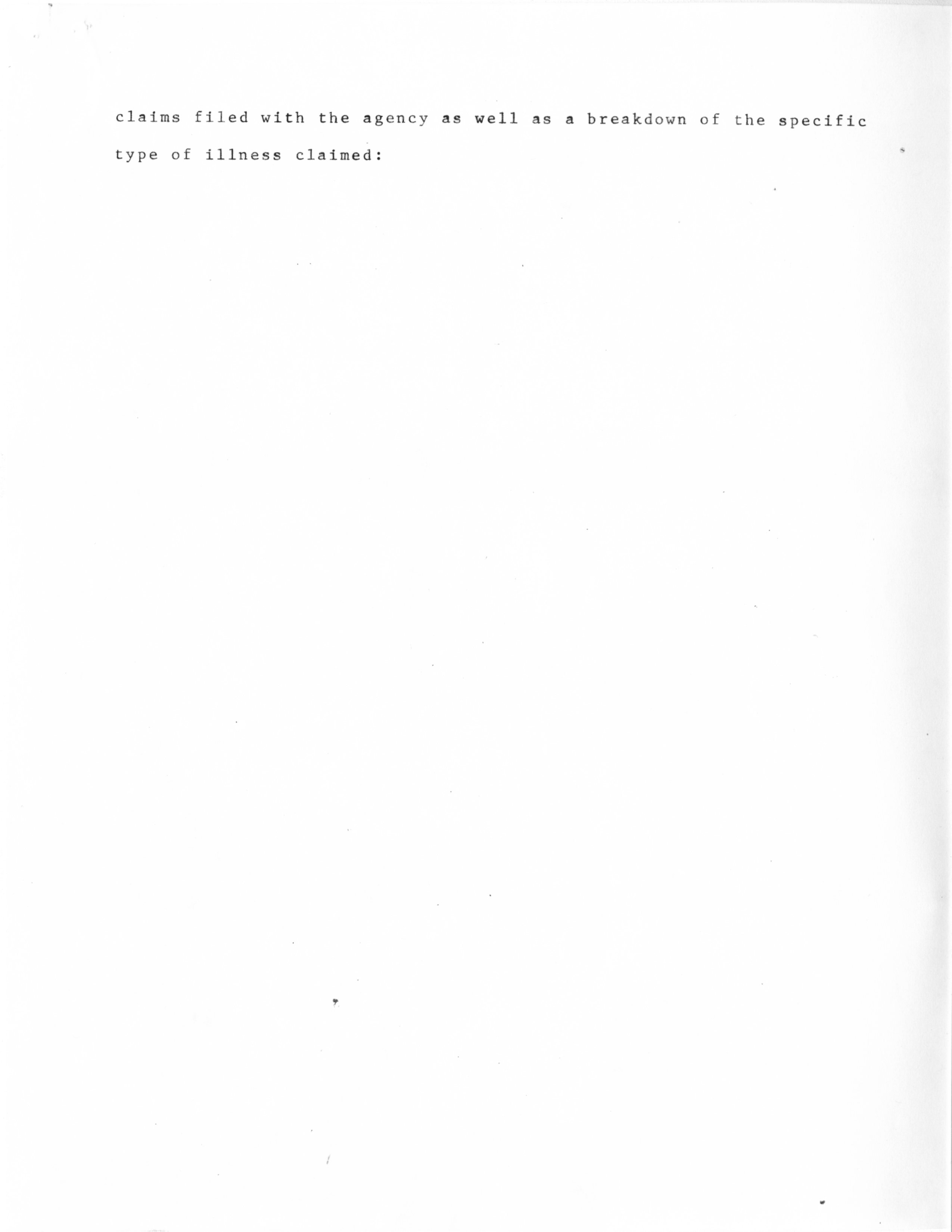}}}
}
\caption{Document examples}
  \label{fig.ex}
\end{figure}

High-quality documents were subjected to automatic spelling correction
via the {\sc Hunspell} package (477 documents or 11\% of the total documents). We assumed the errors in high-quality documents were more correctable than those in mid-quality documents. Correcting mid-quality documents could inflate the noise in these readings if similar strings of noise were corrected to the same words. This left 3,078 remaining documents for analysis. From these documents, we removed stop words from the list provided in the {\sc tidytext} R package \citep{silgeandrobinson16}. ``Stop words'' are words such as ``the,'' ``of,'' and ``is'' that are presumed to contain very little meaning in the context of topic modeling. Additionally, we performed stemming using Porter's algorithm implemented in the {\sc SnowballC} R package \citep{bouchet15}. Stemming aims to combine words with very similar meaning (e.g. ``dance,'' ``dancer'', and ``dances'' would all be stemmed to ``dance''). Figures \ref{fig.ex} presents examples of a low-quality and high-quality document. 

%
%

The output we obtain from the examples of a low- and high- quality document 
\begin{itemize}

\item Low-quality document OCR output: $w A \_. ({\textbf{???}}) Mwmw \_ m
wwwwffl:mmmszflwm\_\_ M\_ W.. : WM m gggfi wwwm:m§;¢fi£\& ;flmjwmiwW:iwm
My flww:fim\_ iw:\_ fiWWW:W Wfi iv fiww: ,1,meJig-fig;Wii:I:ngwwwmflwm: y
\_ g\_ M14\%;\_w;-m\_ mWWWWMTWWWm, \_ f -- WWW W
\# 4:.\_\_\_\_ W \_\_\_ . \_\_ - \_\_\_ r \_ .:n \_ ,\_\_\_ w\_\_\_\_ m\_ ,g,.\_ ,\_\_ w, W\_ W\# Wmia;figfio£e15e$ ...

\item High-quality document OCR output: claims filed with the agency as well as a breakdown of the specific type of illness claimed: `
\end{itemize}

\begin{figure}
\centering
\includegraphics[width=0.85\linewidth]{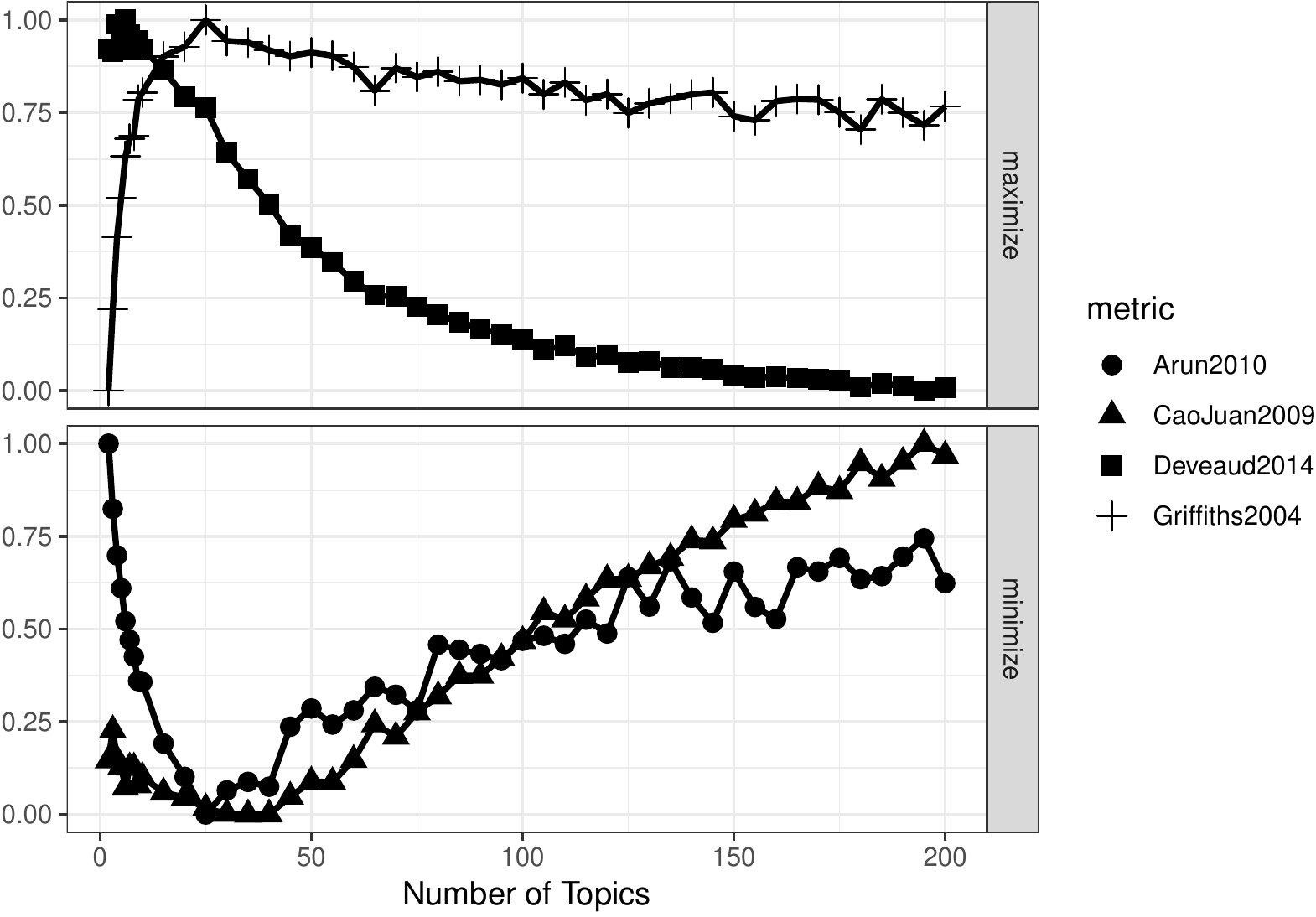}
\caption{\label{fig:FindTopicsNumberPlotPapers}Various metrics used to
choose the optimal number of topics for document collection}
\end{figure}

\begin{figure}
\centering
\includegraphics[width=0.99\linewidth]{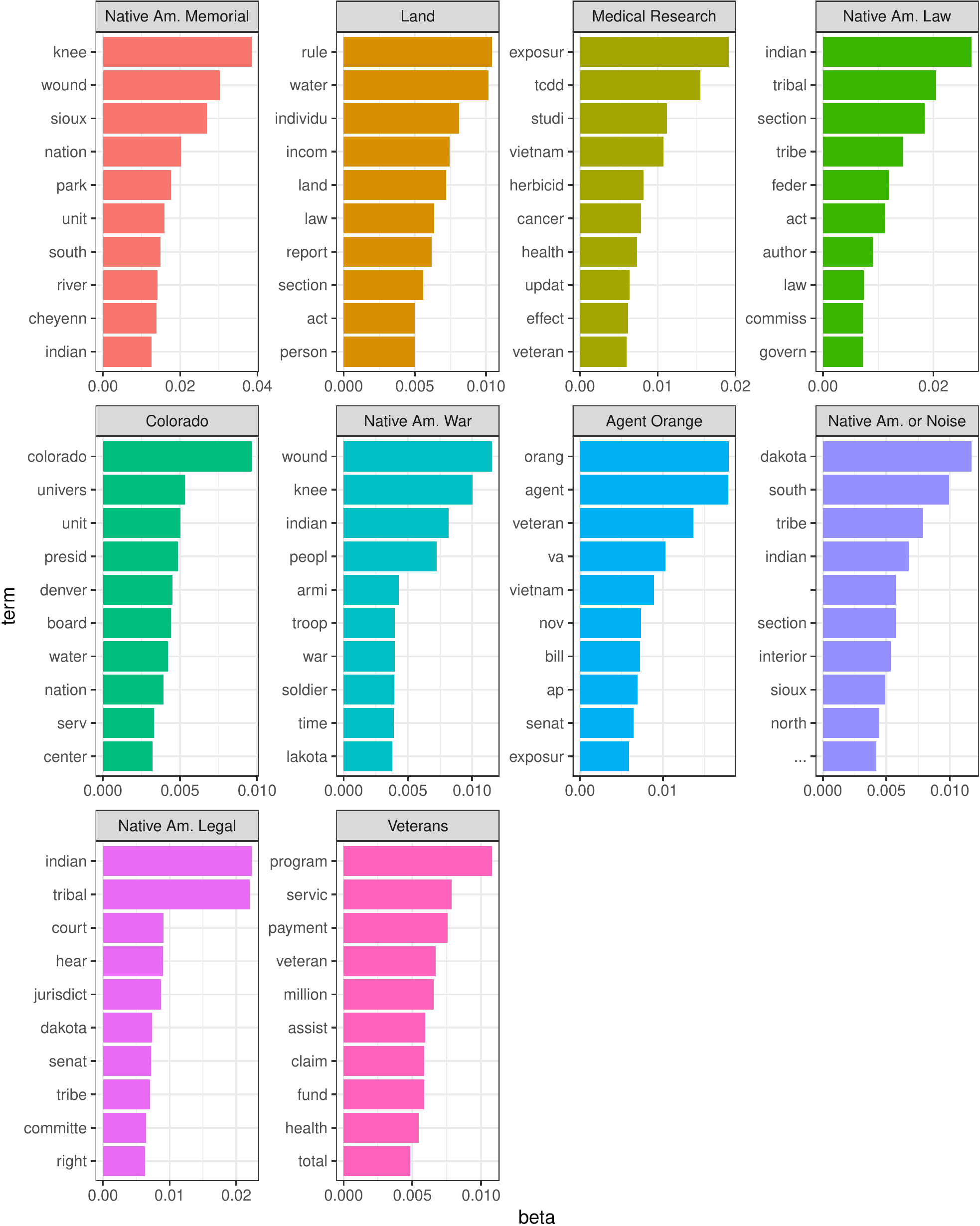}
\caption{\label{fig:PapersTopTermsPlot}Top words for each topic in the paper
documents}
\end{figure}

The LDA topic model is fitted to the the data for different number of topics. To find the optimal number of topics to model in our dataset, we used a variety of metrics as implemented in \cite{nikita16}. The re-scaled results of these are presented in Figure~\ref{fig:FindTopicsNumberPlotPapers}. As a compromise between the
metrics, we initially attempted modeling with 15 topics. However, the
differences between the topics were unclear, with each modeled topic
clearly representing a distinct actual topic, but with some topics
seeming to be duplicated. To eliminate this redundancy, we created a new model with only 10 topics. These topics were largely similar to the
initial 15 but without the redundancy. These 10 topics are used in the
remainder of this analysis. Next, we labeled the resulting 10 topics by hand using the bar plots presented in Figure \ref{fig:PapersTopTermsPlot} and word clouds for
each topic (not shown). Our labels are given in the title of each bar
chart. Beta is \(P(\text{term} \mid \text{topic})\).

\begin{figure}
\centering
\includegraphics[width=0.8\linewidth]{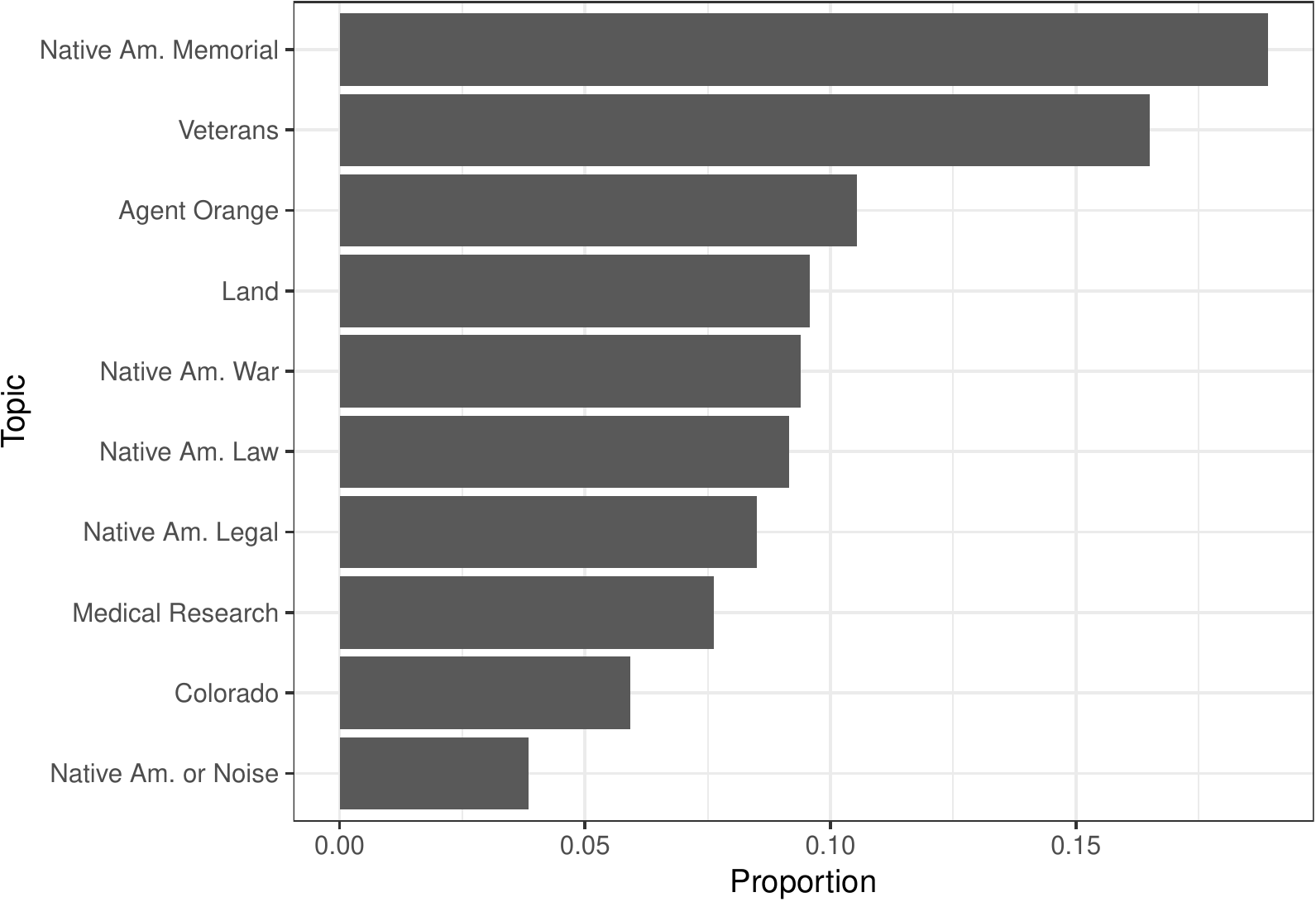}
\caption{\label{fig:PapersTopicPropPlot}Proportion of each topic in the
paper documents}
\end{figure}

We also examined the overall proportion of the topics throughout the
collection. From Figure \ref{fig:PapersTopicPropPlot}, it is clear that
documents about American Indians dominate the digitized collection, with approximately 50\% of all the text coming from a topic related to Native Americans. Because the digitized documents are only a small subset of the total collection, this may not be representative of the contents of the rest of the documents but indicates the interests of researchers on this topic. The second most popular topic in the sample collection was about veterans and ``agent orange". 
As a representative of South Dakota, Daschle was, of course, concerned
with Native American affairs and worked on many related bills. The
effects of Agent Orange on Vietnam veterans was also a topic of
importance to Daschle. The chemical was used in wartime to deforest
areas where the enemy could hide, but had unintentional, long-lasting
effects on soldiers who were exposed to it. Daschle advocated for a bill to provide permanent disability benefits to those who suffered under the chemical's effects \citep{gough03,daschleetal08}.

\subsection{Emails}

\begin{figure}[ht]
\centering
\includegraphics[width=0.55\linewidth]{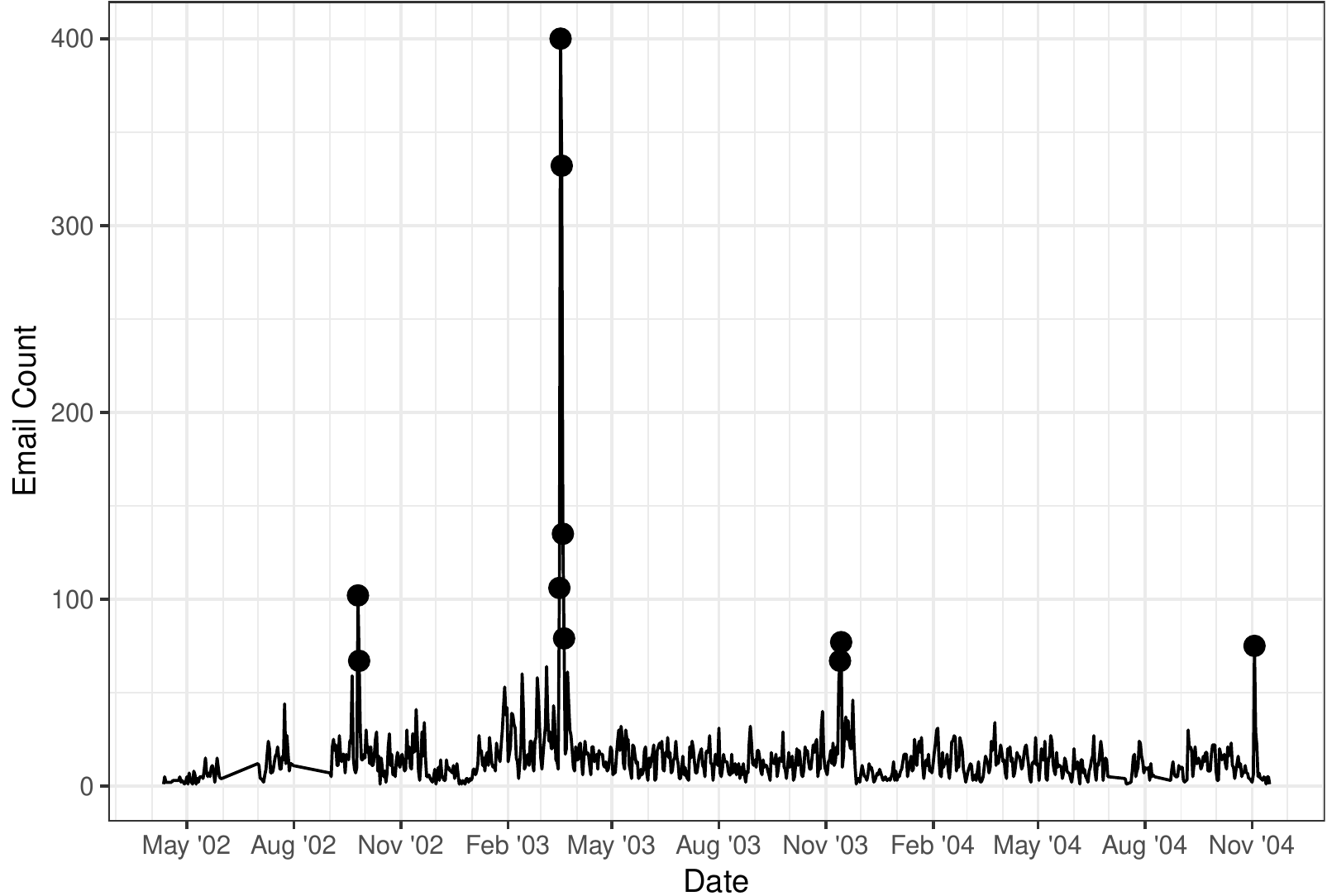}
\caption{\label{fig:EmailsByDatePlot}Number of emails per day in the email
documents}
\end{figure}

While Dashcle served in congress from January 3, 1979 to January 3,
2005, the collection of emails available for analysis spans only the
final years of his career, from April 11, 2002 to November 16, 2004.
These emails range in length from one word to 59,740 words, with 80\% of the emails containing fewer than 260 words. Before looking into the contents of the emails, we first attempt to identify major events in Daschle's career by examining email frequency over time.  The top 10 most active days are highlighted by black dots in
Figure \ref{fig:EmailsByDatePlot}.
\vspace{10pt}

\begin{figure}[H]
\centering
 \mbox{
 \subfigure[CNN:Daschle on `politicizing the war, \url{http://www.cnn.com/2002/ALLPOLITICS/09/25/daschle.comments/}]{\scalebox{0.2}
      {\includegraphics{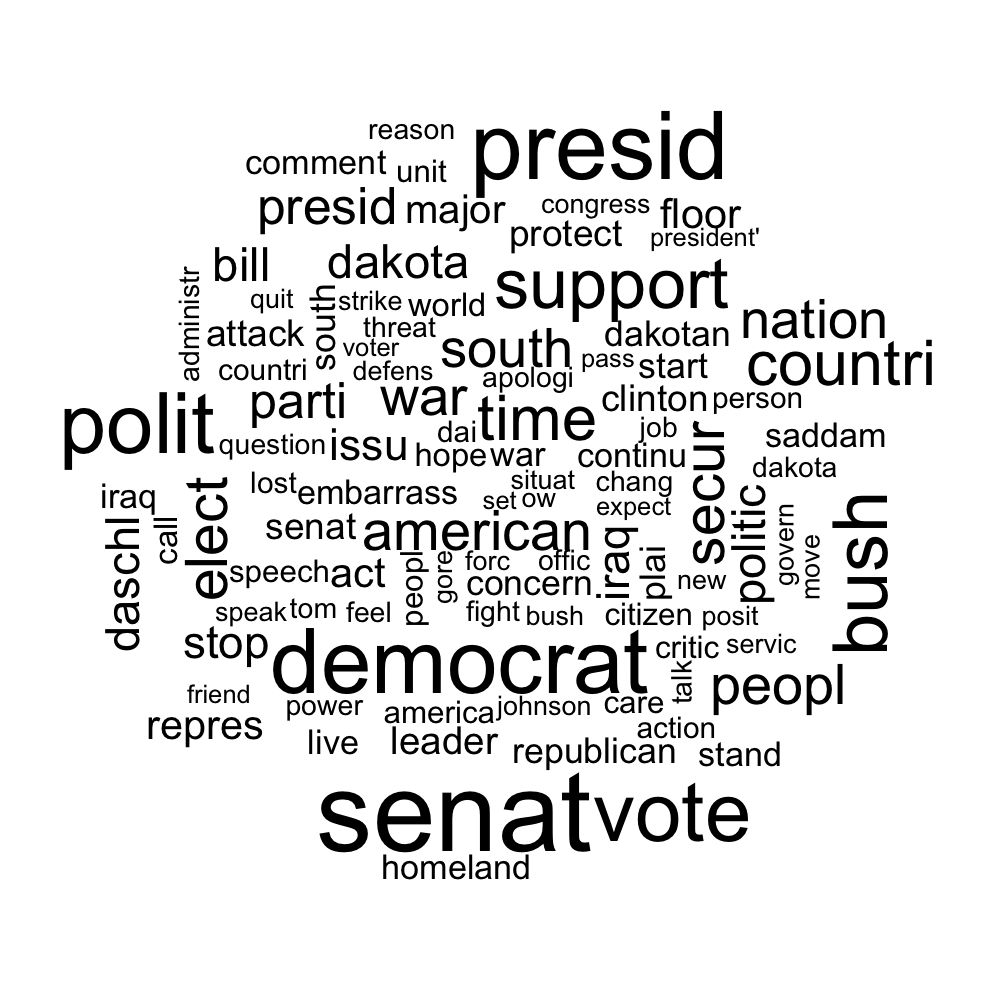}}}
}
\centering
 \mbox{
 \subfigure[USA
TODAY: Daschle: Bush diplomacy fails `miserably', \url{http://usatoday30.usatoday.com/news/washington/2003-03-18-daschle-bush_x.htm}]{\scalebox{0.2}
      {\includegraphics{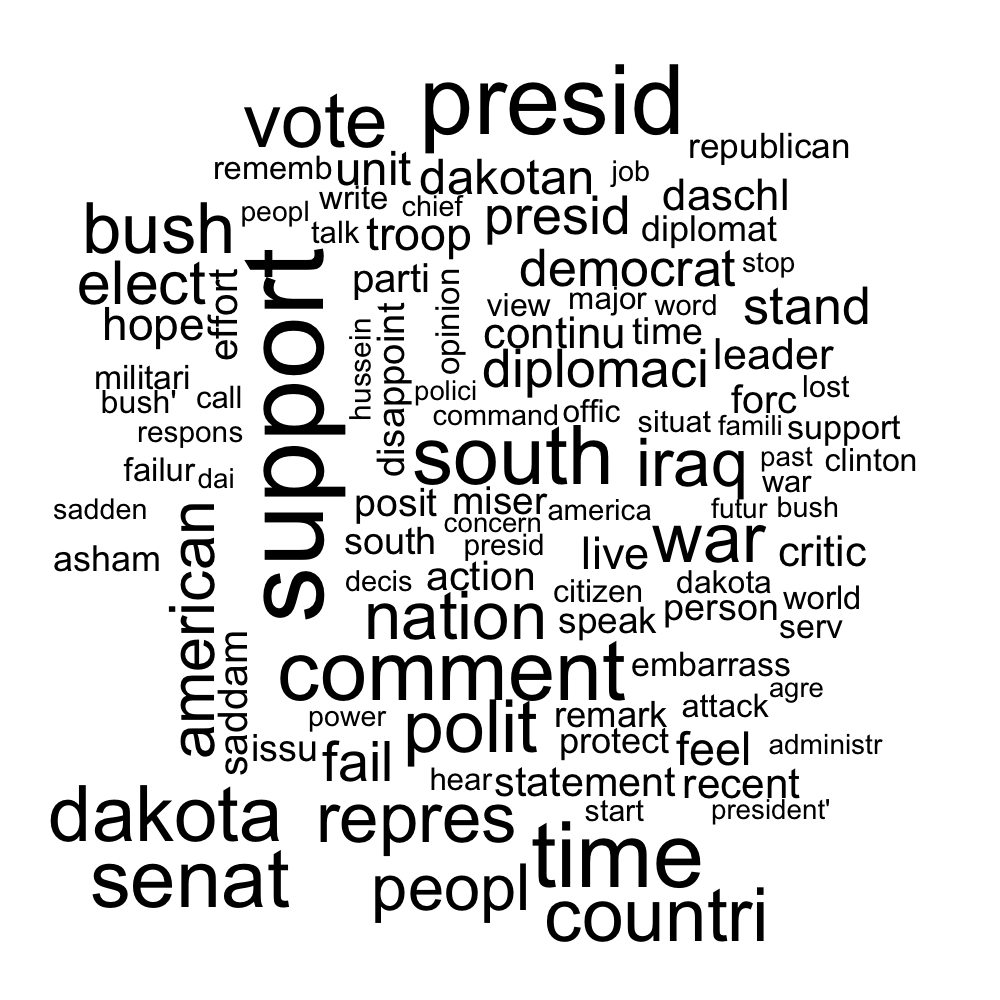}}}
}

\centering
 \mbox{
 \subfigure[CNN:
Democrats begin filibuster against Estrada, \url{http://www.cnn.com/2003/ALLPOLITICS/02/13/senate.estrada/}]{\scalebox{0.2}
      {\includegraphics{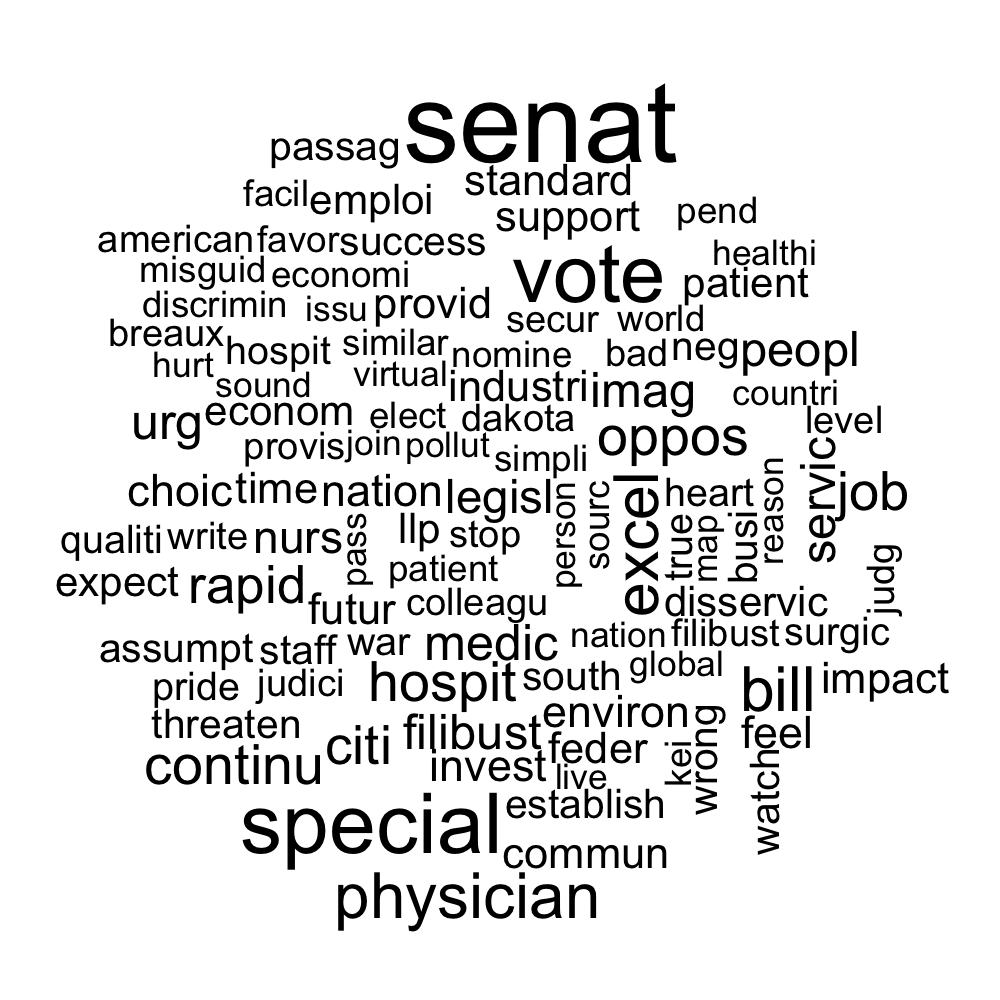}}}
}
 \centering
  \mbox{
       \subfigure[CNN:
Thune unseats Senate minority leader Daschle, \url{http://www.cnn.com/2004/ALLPOLITICS/11/03/senate.southdakota/}]{\scalebox{0.2}
      {\includegraphics{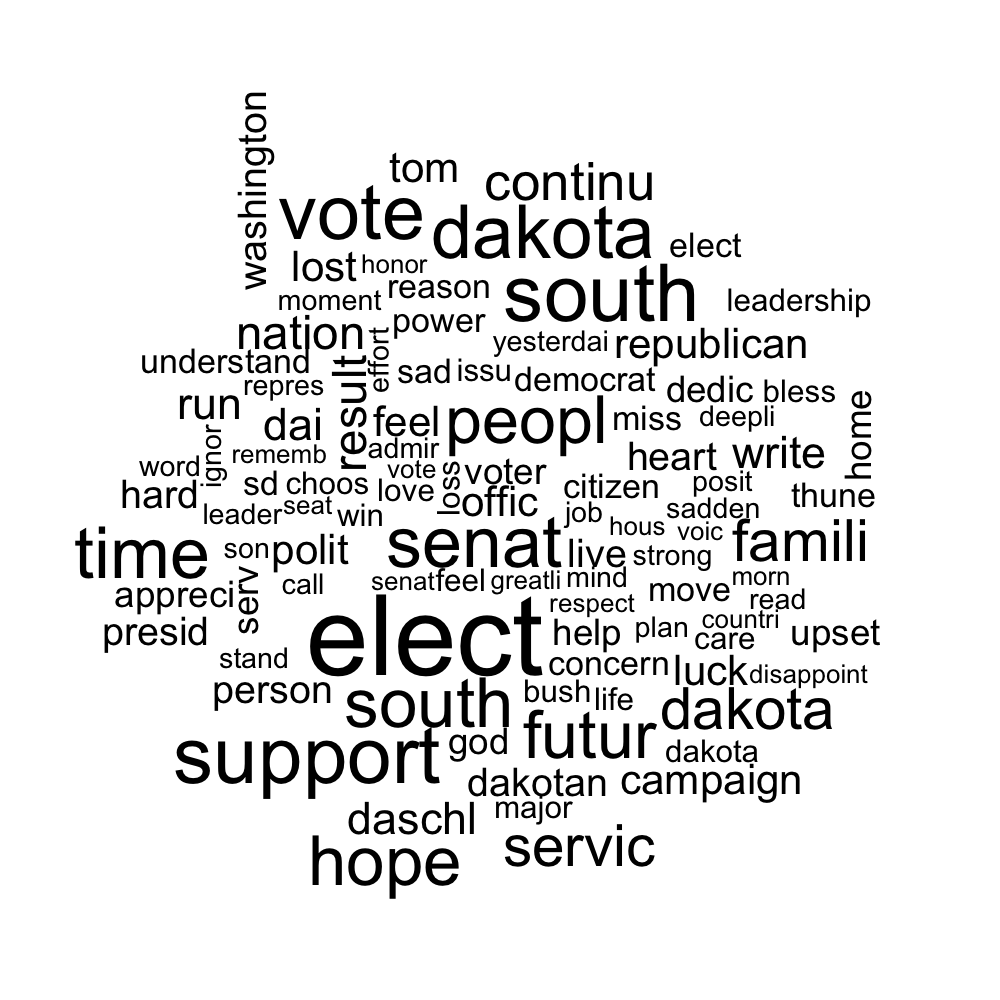}}}
}
\caption{Wordcloud associated with high volume email dates}
  \label{fig.datewordclouds}
\end{figure}

We looked in to the specific dates to explain these dramatic increases in email volume by looking at historical context, as well as the content of the emails themselves. In Figure~\ref{fig.datewordclouds} we present the dates, a headline from a news source which we assume to be the cause of the influx of emails, and a word cloud detailing the contents of the emails from the relevant date range. The content of the emails seem to correspond with the news articles. In two cases where Daschle criticized the Iraq War, constituents were upset and wrote in to say he should ``support'' the war and they were ``disappointed'' with his ``comment.'' The emails in November, 2003 were less focused. Many emailed in support or opposition to the Democrats' decision to block George W. Bush's nomination of Miguel Estrada by using a ``filibuster'' to prevent his confirmation vote from taking place.
Additionally, many emailed on these days about health care, opposing the ``Breaux Amendment'' to the Medicare Prescription Drug, Improvement, and Modernization Act. Emails about the amendment came almost exclusively
from constituents who identified themselves as employees of the Black
Hills Surgery Center in Rapid City who felt that bill would be
disastorous for their institution. The final spike in emails came on the
night Daschle lost his reelection campaign to John Thune in 2004.
Constituents used words like ``vote,'' ``elect,'' and thanked Daschle
for his ``service.'' This is the final day for which emails are provided in the dataset.

\begin{figure}
\centering
\includegraphics[width=0.8\linewidth]{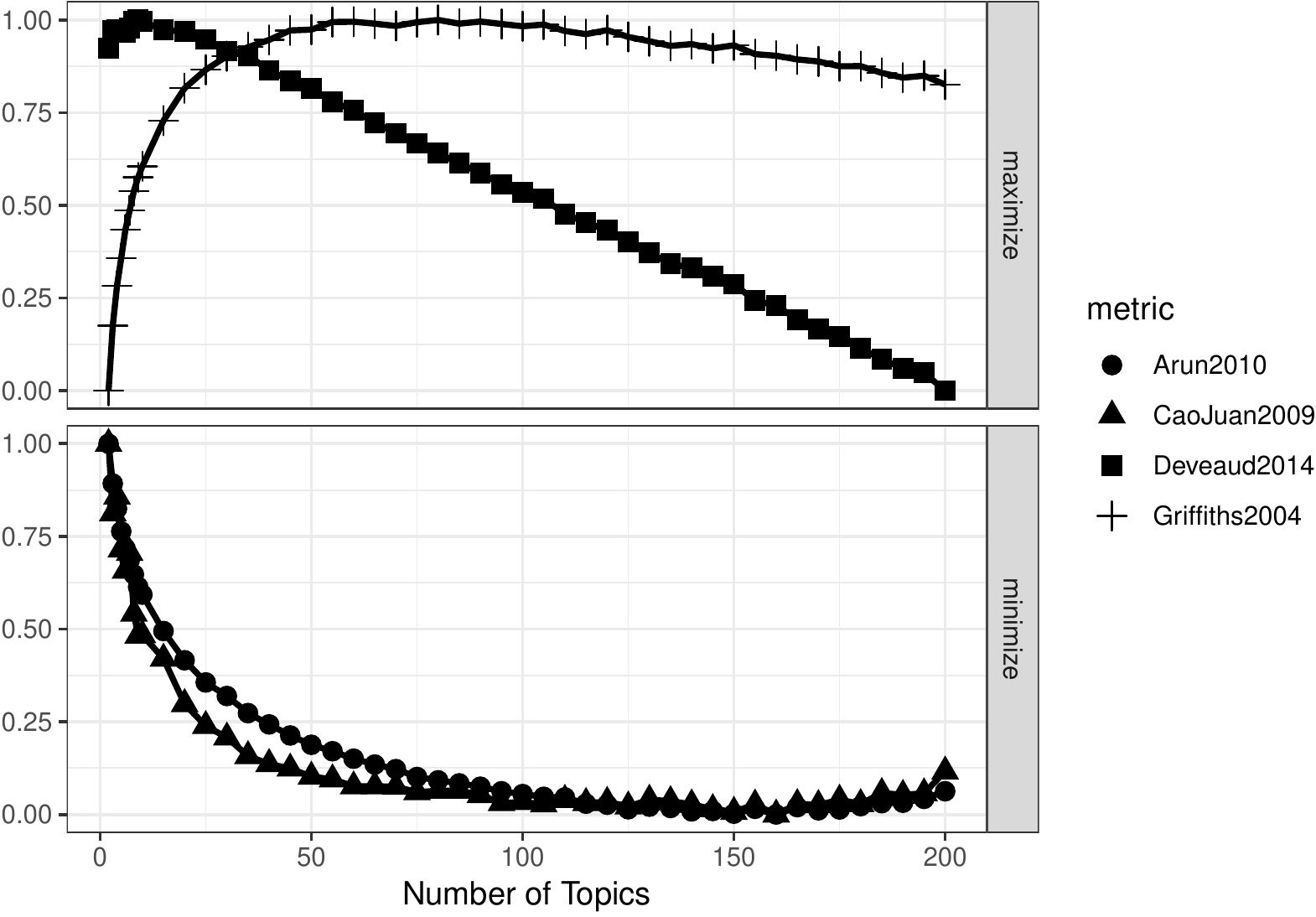}
\caption{\label{fig:EmailsFindTopicsNumberPlot}Various metrics used to
choose the optimal number of topics for the paper documents}
\end{figure}

\begin{figure}
\centering
\includegraphics[width=0.99\linewidth]{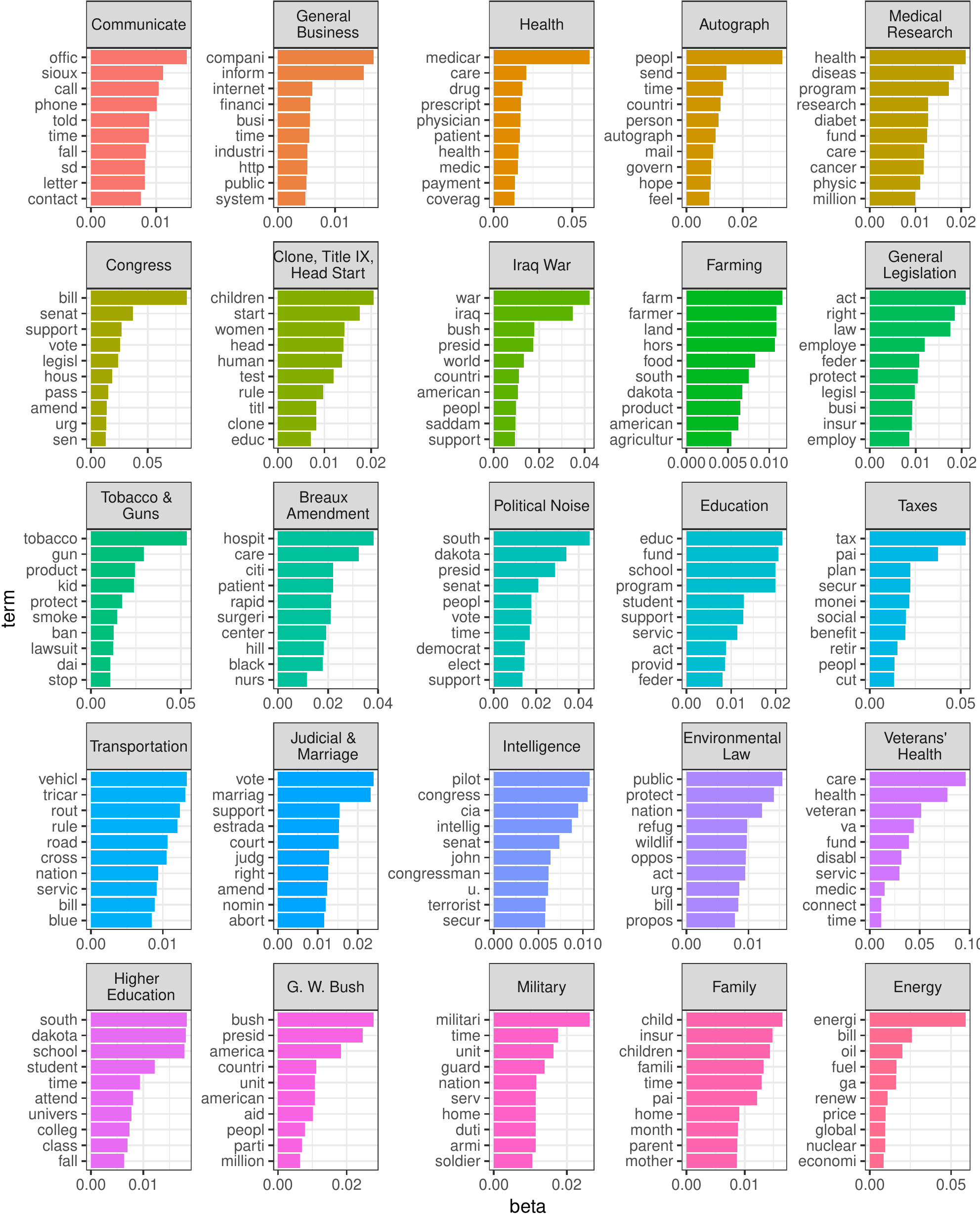}
\caption{\label{fig:emailsTopTermsPlot}Top words for each topic in the email
documents}
\end{figure}

Next, we examine the results of LDA topic modeling on all emails. To
decide on a number of topics to use, we use a variety of metrics,
displayed in Figure \ref{fig:EmailsFindTopicsNumberPlot}. After
initially modeling with 15 topics, we determined that some topics
appeared to be mixtures of two or more other topics. To improve the
separation of topics, we created a new model with 25 topics, which
seemed to separate the topics well without leading to duplication. 
We label the resulting 25 topics by hand using the bar plots presented
in Figure \ref{fig:emailsTopTermsPlot} and word clouds for each topic
(not shown). Our labels are given in the title of each bar chart. Beta
is \(P(\text{term} \mid \text{topic})\).

We also examined the relative quantity of emails with respect to the
topics. Topics reflecting the ``major events'' detailed above, such as
``Iraq War'' and ``Judicial \& Marriage'' are some of the most prominent
in the entire corpus. ``Political Noise,'' the most popular topic is one
of the few without an obviously applicable label. It consists mostly of
words like ``south,'' ``dakota,'' ``democrat,'' ``republican,'' and
``vote.'' We conjecture that this topic includes political words used in
combination with a variety of issues. For example, a document discussing
the Democrats' stance on an education bill would be a mix of ``Political
Noise'' and ``Education'' and an email discussing farmers in South
Dakota would be a mix of ``Political Noise'' and ``Farming.''

\begin{figure}
\centering
\includegraphics[width=0.8\linewidth]{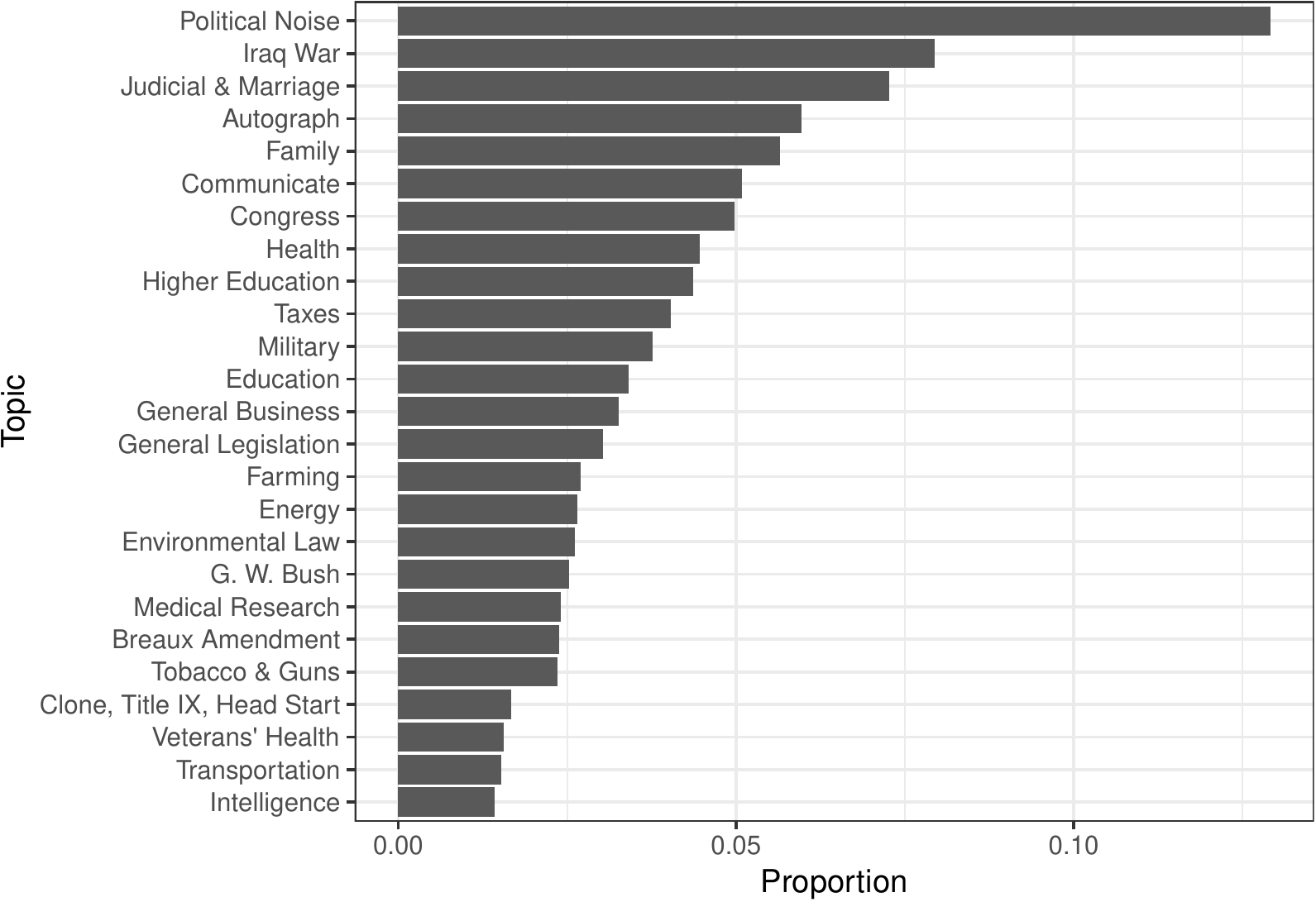}
\caption{\label{fig:unnamed-chunk-3}Proportion of each topic in the email
documents}
\end{figure}

\begin{figure}[H]

{\centering \includegraphics[width=0.8\linewidth]{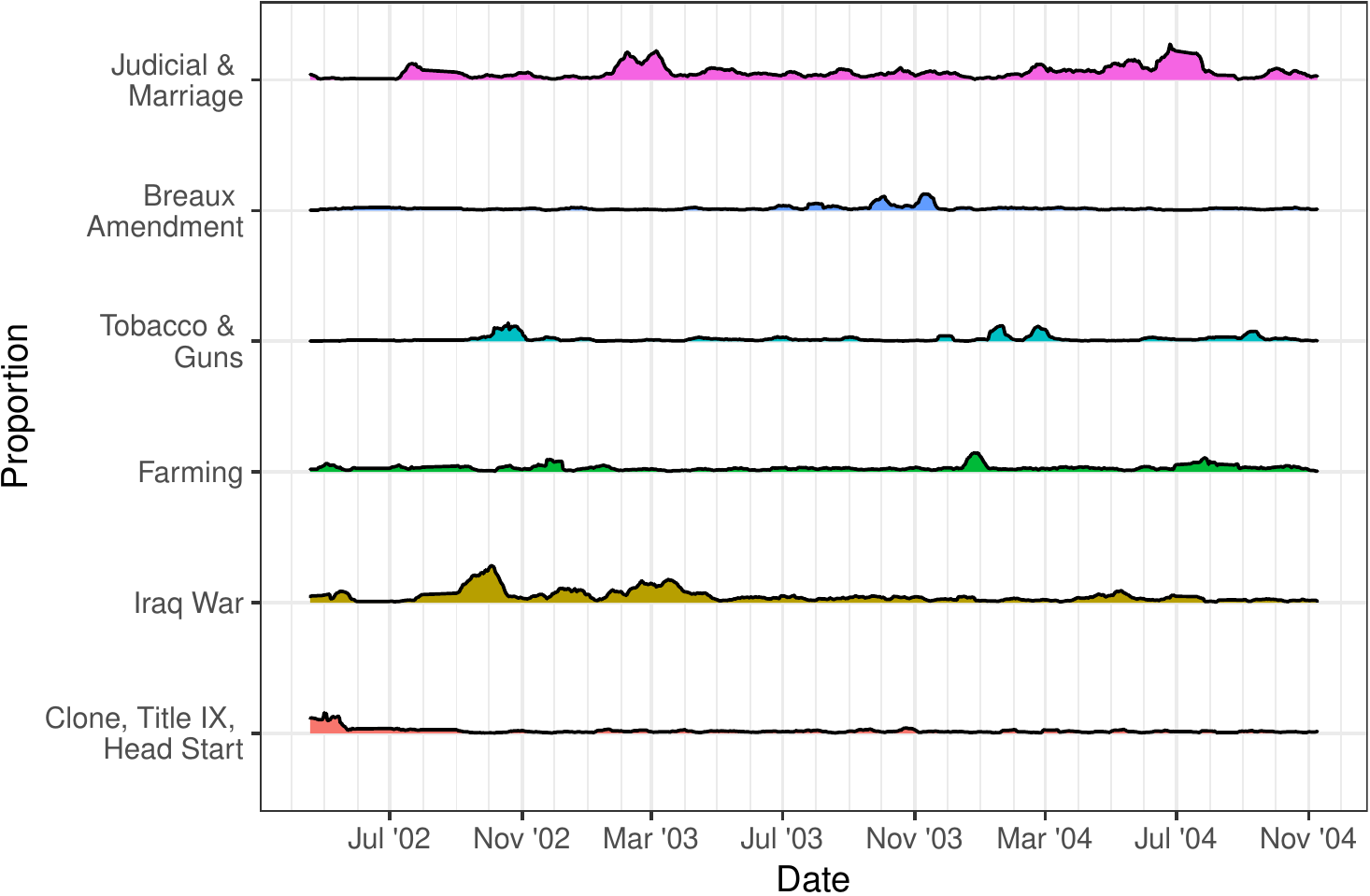} 

}

\caption{Selected topic prevalence over time}\label{fig:topicsOverTimePlot}
\end{figure}

Next we examine how the content of emails changed over time. Topics not
presented in Figure \ref{fig:topicsOverTimePlot} showed consistent
presence over time. Of the topics with visible trends, most correspond
with relevant events in history. The increase in ``G. W. Bush'' emails
corresponds to Daschle's controversial comments discussed earlier in
this section. The spike in ``Judicial \& Marriage'' emails relates to
the nomination of Miguel Estrada in 2002, as well as the Federal
Marriage Amendment, a bill that defined marriage in the United States as
being between a man and a woman, in 2004. Emails involving high
proportions of Political Noise peaked when Daschle lost his seat to
Thune in the November 2004 election. The increase in Tobacco \& Guns
emails in late 2002 appears to be related to a very passionate
individual or group who emailed Daschle the same message about Tobacco
and marketing to children almost daily for several weeks. These messages
persist over time, but were particularly concentrated in late 2002.
Similarly, one or more people repeatedly sent a message about gun laws
in mid 2004. The prevalence of messages about the Iraq War is
commensurate with the the war's beginning. Messages regarding Clone,
Title IX, Head Start appear most prevalently in 2002, when the Human
Cloning Prohibition Act of 2001 was being considered by congress.

\section{Discussion}
\label{sect.disc}
Topic modeling on the Dashcle collection appears to be a successful
method of summarizing the concerns of Daschle and his constituents while
maintaining the privacy of South Dakotans. In general, the topic models
and other explorations revealed patterns which might be expected to be
present in the data. Topics from the paper documents were dominated by
the massive volumes of research on specific topics that were scanned to
be used in further research. Email volume increased following
controversial or surprising events, but was generally consistent in
volume, as well as topic, focusing on war, family, as well as asking for
autographs. 

In addition to the contribution of converting the digital image type documents to text form, several patterns and topics are uncovered throughout this project. Exploration of the currently digitized portion of the Daschle Collection shows a sufficient quality and quantity of data available. The results identified prominent topics such as healthcare reform, veterans' benefits and others that Daschle was known for during his time in office but will also expose less obvious themes, which may not have been identified by previous researchers. Consequently, the results of this project provides researchers documents that are ready for further analysis and interpretation.

While our analysis is interesting in isolation own, we
believe many future reports could be aided by using our topic models to
uncover information about specific events or concerns of constituents
during Daschle's tenure in Congress.

\bibliographystyle{plainnat}
\bibliography{references}

\begin{thebibliography}{22}
\providecommand{\natexlab}[1]{#1}
\providecommand{\url}[1]{\texttt{#1}}
\expandafter\ifx\csname urlstyle\endcsname\relax
  \providecommand{\doi}[1]{doi: #1}\else
  \providecommand{\doi}{doi: \begingroup \urlstyle{rm}\Url}\fi

\bibitem[Arun et~al.(2010)Arun, Suresh, Madhavan, and Murthy]{arunetal10}
Rajkumar Arun, Venkatasubramaniyan Suresh, CE~Veni Madhavan, and MN~Narasimha
  Murthy.
\newblock On finding the natural number of topics with latent dirichlet
  allocation: Some observations.
\newblock In \emph{Pacific-Asia conference on knowledge discovery and data
  mining}, pages 391--402. Springer, 2010.

\bibitem[Blei et~al.(2003)Blei, Ng, and Jordan]{bleietal03}
D.~Blei, A.~Ng, and M.~Jordan.
\newblock Latent dirichlet allocation.
\newblock \emph{Journal of Machine Learning Research}, 3:\penalty0 993--1022,
  2003.

\bibitem[Blevins(2011)]{blevins11}
C~Blevins.
\newblock Topic modeling historical sources: Analyzing the diary of martha
  ballard.
\newblock \emph{Proceedings of Digital Humanities. Stanford, CA: Digital
  Humanities}, 2011.

\bibitem[Bouchet-Valat(2015)]{bouchet15}
Milan Bouchet-Valat.
\newblock Snowballc: Snowball stemmers based on c libstemmer utf-8 library.
\newblock https://CRAN.R-project.org/package=SnowballC, 2015.

\bibitem[Buurma(2015)]{buurma15}
Rachel~Sagner Buurma.
\newblock The fictionality of topic modeling: Machine reading anthony
  trollope's barsetshire series.
\newblock \emph{Big Data \& Society}, 2\penalty0 (2), 2015.

\bibitem[Cao et~al.(2009)Cao, Xia, Li, Zhang, and Tang]{caoetal09}
Juan Cao, Tian Xia, Jintao Li, Yongdong Zhang, and Sheng Tang.
\newblock A density-based method for adaptive lda model selection.
\newblock \emph{Neurocomputing}, 72\penalty0 (7-9):\penalty0 1775--1781, 2009.

\bibitem[Chaudhuri et~al.(2017)Chaudhuri, Mandaviya, Badelia, and
  Ghosh]{chaudhurietal17}
Arindam Chaudhuri, Krupa Mandaviya, Pratixa Badelia, and Soumya~K Ghosh.
\newblock Optical character recognition systems for english language.
\newblock In \emph{Optical Character Recognition Systems for Different
  Languages with Soft Computing}, pages 85--107. Springer, 2017.

\bibitem[Daschle(2017)]{daschle17}
T.~Daschle.
\newblock Senator thomas a. daschle congressional research.
\newblock \url{https://www.sdstate.edu/daschle-study}, 2017.
\newblock Accessed: 2017-09-18.

\bibitem[Daschle et~al.(2008)Daschle, Greenberger, and Lambrew]{daschleetal08}
Tom Daschle, Scott~S Greenberger, and Jeanne~M Lambrew.
\newblock \emph{Critical: what we can do about the health-care crisis}.
\newblock Macmillan, 2008.

\bibitem[Deveaud et~al.(2014)Deveaud, SanJuan, and Bellot]{deveaudetal14}
Romain Deveaud, Eric SanJuan, and Patrice Bellot.
\newblock Accurate and effective latent concept modeling for ad hoc information
  retrieval.
\newblock \emph{Document num{\'e}rique}, 17\penalty0 (1):\penalty0 61--84,
  2014.

\bibitem[Gough(2003)]{gough03}
MICHAEL Gough.
\newblock The political science of agent orange and dioxin.
\newblock \emph{Politicizing Science: The Alchemy of Policymaking}, pages
  193--225, 2003.

\bibitem[Griffiths and Steyvers(2004)]{griffithsandsteyvers04}
Thomas~L Griffiths and Mark Steyvers.
\newblock Finding scientific topics.
\newblock \emph{Proceedings of the National academy of Sciences}, 101\penalty0
  (suppl 1):\penalty0 5228--5235, 2004.

\bibitem[Ian(2014)]{ian14}
Fellows Ian.
\newblock wordcloud: Word clouds. r package version 2.5, 2014.

\bibitem[Jurafsky and Martin(2009)]{jurafskyandmartin09}
Dan Jurafsky and James~H Martin.
\newblock Speech and language processing: An introduction to natural language
  processing, computational linguistics, and speech recognition, 2009.

\bibitem[Meyer et~al.(2008)Meyer, Hornik, and Feinerer]{meyeretal08}
David Meyer, Kurt Hornik, and Ingo Feinerer.
\newblock Text mining infrastructure in r.
\newblock \emph{Journal of statistical software}, 25\penalty0 (5):\penalty0
  1--54, 2008.

\bibitem[Nelson(2011)]{nelson11}
Robert~K Nelson.
\newblock Of monsters, men—and topic modeling.
\newblock \emph{New York Times (May 29, 2011)}, 2011.

\bibitem[Nemeth(Accessed June 2017)]{hunspell19}
Laszlo Nemeth.
\newblock Hunspell.
\newblock "http://hunspell.github.io", Accessed June 2017.

\bibitem[Nikita(2016)]{nikita16}
Murzintcev Nikita.
\newblock ldatuning: Tuning of the latent dirichlet allocation models
  parameters.
\newblock \emph{R package version 0.2-0, URL https://CRAN. R-project.
  org/package= ldatuning}, 2016.

\bibitem[Ooms(2017{\natexlab{a}})]{ooms17a}
Jeroen Ooms.
\newblock tesseract: Open source ocr engine for r.
\newblock \emph{R package version}, 1, 2017{\natexlab{a}}.

\bibitem[Ooms(2017{\natexlab{b}})]{ooms17b}
Jeroen Ooms.
\newblock Hunspell: High-performance stemmer, tokenizer, and spell checker for
  r. r package version 2.3.
\newblock \emph{URL: https://CRAN. R-project. org/package= hunspell},
  2017{\natexlab{b}}.

\bibitem[{R Development Core Team}(2016)]{R16}
{R Development Core Team}.
\newblock \emph{R: A Language and Environment for Statistical Computing}.
\newblock R Foundation for Statistical Computing, Vienna, Austria, 2016.
\newblock URL \url{http://www.R-project.org}.

\bibitem[Silge and Robinson(2016)]{silgeandrobinson16}
Julia Silge and David Robinson.
\newblock tidytext: Text mining and analysis using tidy data principles in r.
\newblock \emph{The Journal of Open Source Software}, 1\penalty0 (3):\penalty0
  37, 2016.

\end{thebibliography}

\end{document}